\documentclass{article}
\usepackage{graphicx}

\begin{document}                                 
\noindent {\bf Non-extensivity in magnetic systems: possible impact on Mossbauer results}
\vskip1.5cm
\noindent {\bf Ashok Razdan}

\noindent {\bf Astrophysical Sciences Division}

\noindent {\bf Bhabha Atomic Research Centre}

\noindent {\bf Trombay, Mumbai- 400085 }
\vskip 0.2cm
email:akrazdan@barc.gov.in
\vskip 0.1cm
Tel: +91 22 25591798; fax: +91 22 25505151
\noindent {\bf Abstract :}

Energy positions of pure magnetic transitions 
in Mossbauer Spectroscopy are calculated  using non-extensive  approach.
It is observed that these new calculated energy  positions so obtained, may have strong overlap with   
those energy positions obtained from  combined effect of magnetic and quadrupole interactions
using standard statistical physics.

\noindent {\bf Motivation:}

Non-extensive statistics is being increasingly used to explain
anomalous results  observed in  various physical 
systems like 
turbulence in plasma,Cosmic ray background radiation, self gravitating systems,
econo-physics,electron positron annihilation, chaos, linear response 
theory , Levy type anomalous super diffusion ,  
Lamb Mossbauer factor,specific heat in glasses ,low dimensional systems [1-9] etc. 
The non-extensive approach is based on non-extensive entropy which is given as
\begin{equation}
S_q  = \frac{ 1- \sum p_i^{q}}{q -1}
\end{equation}
$p_i$ are the probabilities of the microscopic states with $\sum p_i$=1.
In the  limit of q $\rightarrow$ 1, $S_1$= -$\sum p_i ln(p_i)$ which is
Boltzmann-Gibbs-Shannon entropy.
It has been shown that non-extensive features get manifested in those systems 
which have long range forces, long memory effects,inhomogeneous systems or in those systems which 
evolve in (non Euclidean like space-time) fractal space time [10 and reference therein]. 
It has been shown in recent times that non-extensivity is also relevant to
magnetic systems.
Using inputs from various experimental studies,
manganites have been identified as possible non-extensive objects.
Manganites have long range Coulomb interactions[ 11,12,13], fractal like clusters [14,15]
and intrinsic inhomogeneity [16,17,18]
It has been
further shown that non-extensivity parameter 'q' is a measure of inhomogeneity .
Correct predictions about bulk magnetization have been made using non-extensive approach [19,20,21].
Non-extensivity in griffths phase has also been investigated [22,23]. 
This relevance of non-extensivity in magnetism [ 24] has motivated us to explore
its impact on Mossbauer parameters  of magnetic systems.

\noindent {\bf Nuclear Zeeman splitting:}

Magnetic hyperfine splitting arises from interaction between magnetic moments of the
ground and excited states of nucleus [25] with the internal or external magnetic field.
The interaction Hamiltonian is given as 
\begin{equation}
H= -\mu. H_n 
\end{equation}
where $\mu$ is magnetic  moment and $H_n$ is the magnetic field.
This interaction completely lifts the degeneracy of nuclear levels of spin I and
energy $E_0$ is split into (2I+1) levels.
The energy levels obtained are 
\begin{equation}
E_m=E_0 - g_n \mu_n H_n m_I
\end{equation}
where $m_I$= I,I+1,.. -I, $g_n$ is the splitting factor ( gryomagnetic ratio), $\mu_n$ is 
the nuclear magneton.
The magnetic interaction splits both the ground and excited states between which transitions
take place. 
Thus in  $^{57} Fe $ six transitions are obtained 
and  the transition  probabilities are given by well know Clebsch-Gordon coefficients.
The transitions probabilities, their angular variations and  energy positions  are given in reference [25].

For $\pm \frac{3}{2} \rightarrow \pm \frac{1}{2}$  transition, the energy
positions of Ist and 6th transition corresponds to [25]. 
\begin{equation}
Energy-Position (1,6)=(E_0 \pm \frac{3}{2} g_e \mu_n H_n \pm \frac{1}{2} g_g \mu_n H_n)
\end{equation}
where $g_e$ and $g_g$ are the ratios  of the nuclear magnetic moment to the
nuclear magneton $\mu_n$ for the ground and the excited states respectively.
 
\noindent {\bf Non-extensive approach in magnetic splitting:}

For non-extensive case, [19,20,21,22] interaction Hamiltonian is given by  
\begin{equation}
H=-\mu_{ne}. H_n
\end{equation}
where $\mu_{ne}$ is non-extensive magnetic moment.
The expression for non-extensive magnetization is given as
\begin{equation}
M_q =\frac {\mu_{ne}}{(2-q)}  [ coth_q(x) -\frac{1}{x}]
\end{equation}
where x=$\frac{\mu_{ne} H}{kT}$.
It has been found that  relationship
\begin{equation}
\mu_{ne}= (2-q) \mu 
\end{equation}
hold for 0 $\le q \le 1 $.
This relationship has also been derived using [21] generalized Brillouin function.
To obtain energy position for the non-extensive case, replace $\mu_n$ in equation (4) by $\mu_{ne}$ and use equation
(7).Thus
for $\pm \frac{3}{2} \rightarrow \pm \frac{1}{2}$  transition of $^{57} Fe$ nucleus , non-extensive energy
positions of Ist and 6th transition correspond to 
\begin{equation}
Energy-Position (1,6)=(E_0 \pm \frac{3}{2} (2-q) g_e \mu_n H_n \pm \frac{1}{2} (2-q) g_g \mu_n H_n)
\end{equation}
Similarly for $\pm \frac{1}{2} \rightarrow \pm \frac{1}{2}$
transition of $^{57}Fe$ nucleus , non-extensive energy
positions of 2nd and 5th transition correspond to
\begin{equation}
Energy-Position (2,5)=(E_0 \pm \frac{1}{2} (2-q)g_e \mu_n H_n \pm \frac{1}{2} (2-q) g_g \mu_n H_n)
\end{equation}
For $\mp \frac{1}{2} \rightarrow \pm \frac{1}{2}$
transition of $^{57} Fe$ nucleus , non-extensive energy
positions of 3rd and 4th transition  correspond to
\begin{equation}
Energy-Position (3,4)=(E_0 \mp \frac{1}{2} (2-q)g_e \mu_n H_n \pm \frac{1}{2} (2-q) g_g \mu_n H_n).
\end{equation}
Each energy position in equation (7-9) corresponds to two energies.
Thus the relationship between non-extensive and standard energy position can be written as
\begin{equation}
(Energy-Position)_{non-extensive} \propto (2-q)(Energy-Position)_{standard}
\end{equation}
All the new energy positions obtained from non-extensive  approach  will have q dependence. 

Equations (8-10) have been obtained using non-extensive approach.
In the limit of q $\rightarrow$ 1, non-extensive approach
reduces to standard Boltzmann-Gibbs approach
(hereafter referred as standard case in the text).

\begin{figure}
\includegraphics[angle=270,width=7.0cm]{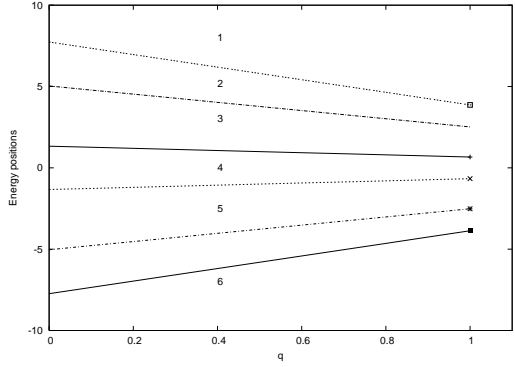}
\caption{ Energy postion v/s q parameter for pure Mossbauer magnetic transitions}
\end{figure}

\noindent{\bf Results and Discussion:}

Energy position dependence on  q parameter has been plotted in figure 1. For q=1  the standard case, the energy positions correspond
to well known positions of normal magnetic splitting. These positions in figure 1 are represented by various symbols
like open/close square, cross, plus etc. in the figure. For all other values of q the energy positions 
in the figure 1 corresponds to non-extensive nature.
In figure 1 curves '1' and '6' correspond to the
$\pm \frac{3}{2} \rightarrow \pm \frac{1}{2}$ transition, curves '2' and '5' corresponds to
$\pm \frac{1}{2} \rightarrow \pm \frac{1}{2}$ transition  and curves '3' and '4' corresponds to
$\mp \frac{1}{2} \rightarrow \pm \frac{1}{2}$ transition respectively.

For pure magnetic splitting in the standard case the positions of energies are fixed.
But combined effect of magnetic and
quadrupole interaction ( for standard case) results in shifting of these energy positions.
The magnetic energy levels [25,26] for the standard case obtained when treating the quadrupole interaction as a perturbation
is given as
\begin{equation}
E= -g_n  \mu_n H_n m_I + ((-1)^{ m_I+ \frac{1}{2}})   \frac{e^2 V_{zz} Q}{4}  ( \frac{3cos^2 \theta -1}{2})
\end{equation}
where Ze is the nuclear charge, eQ is the quadrupole moment and $V_{zz}$ is the electric field gradient along
z axis.
The above equation holds for $e^2 V_{zz} Q << \mu_n H_n$ i.e. quadrupole interaction is much
less than the magnetic interaction.
The combined effect of magnetic and quadrupole interaction in the standard case [25] is parameterized by $\lambda$ which is defined as
\begin{equation}
\lambda = \frac{ \frac{e^2 V_{zz} Q}{2I(2I-1)}}{ \frac{\mu_n H_n}{I}}
\end{equation}
The  value of $\lambda$  is between 0 to 1. In the figure 2 we have plotted energy
position dependence on $\lambda$. The curves  1 and 2 correspond to 
$\frac{3}{2} \rightarrow  \frac{1}{2}$ transition. The curves 3 and 4 correspond to
$ \frac{1}{2} \rightarrow \frac{1}{2}$ transition  and curves 5 and 6 correspond to
$\frac{1}{2} \rightarrow  \frac{1}{2}$ transition respectively.
These curves in figure 2 have been obtained for $\theta$= 0 degree (curves 2,4 and 6) and $\theta$= 90 degrees.
(curves 1,3 and 5) respectively.

For standard case, the presence of quadrupole interaction  affects magnetic 
transitions in such a  way that energy positions are shifted to any position in figure 2 for a given value
of $\lambda$. However,it is  clear from figure 1 that in the non-extensive case, the energy positions are 
shifted even for pure magnetic splitting for a given value of q . Comparing figures 1 and
2, it is obvious  that some of the values may overlap or some curves may even cross each other, in which case
a non-extensive pure magnetic transition may have same energy position  as standard  mixed transition case.

\begin{figure}
\begin{center}
\includegraphics[angle=270,width=7.0cm]{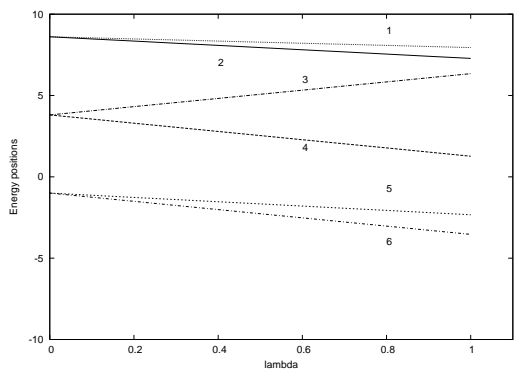}
\caption{Energy Position of Mixed transitions v/s $\lambda$ plot for angle 0 and 90. }
\end{center}
\end{figure}

Large number of Mossbauer results for magnetism [26-29] in general and manganites in particular have been produced
in which presence of quadrupole interaction have been assumed to explain the shift in energy positions
of various transitions. If manganites /magnetic systems  have non-extensive nature than energy positions
will indeed shift, 
yet transitions may be pure magnetic in nature. This means all the calculations of assuming mixed transitions 
for manganites/magnetic systems  may be wrong and need to be reinvestigated.

Thus it is  suggested that for magnetic systems in general and manganites
in particular inverse susceptibility experiments [22] may be used first to establish if the system (sample) is
non-extensive or normal.
For normal systems standard protocol of Mossbauer Spectroscopy should be followed.
For non-extensve systems, using the value of q (obtained from inverse susceptibility experiments) correction should first
be introduced on the energy positions so that non-extensive  energy positions are renormalized to normal system values.
After introducing non-extensive corrections, standard Mossbauer protocol of calculating hyperfine parameters
may  be followed.

\noindent {\bf References: }
\begin{enumerate}
\item B.M.Boghosian, Phys. Rev. E 53(1996)4745
\item C.Tsallis, F.C. Sa Barreto and E.D.Loh, Phys. Rev. E 52(1995) 1447
\item V.H.Hamity and D.E.Barraco, Phys. Rev. Lett. 76(1996)4664
\item I.Bediaga, E.M.F.Curado and J. Miranda, Physica A 286(2000)156
\item C.Tsallis, A.R.Plastino and W. -M.Zheng, Choas,Solitons and Fractals 8(1997)885,\\
      Y.Weinstein, S.Lloyd and C.Tsallis, Phys. Rev. Lett. 89(2002)214101
\item A.Razdan, Phys. Lett. A 321(2004)190
\item A.Razdan, Phys. Lett. A 341(2005)504
\item C.Tsallis, S.V.F. Levy, A.M.C.Souza and R.Maynard, Phys. Rev. Lett. 75(1995)3589
\item I.Koponen, Phys. Rev. E 55(1997)7759
\item M.L.Lyra and C.Tsallis, Phys. Rev. Lett. 80(1998)53
\item C.Tsallis, Physica A 221(1995)277-290,\\  
      C.Tsallis, R.S.Mendes, A.R.Plastino, Physica A 261(1998)534
\item J.Lorenzana et. al., Phys. Rev. B 64(2001)235127
\item J.Lorenzana et. al., Phys. Rev. B 64(2001)235128
\item A.Moreo et.al., Science 283(1999)2034
\item E.Dagotto et. al, Phys. Rep. 344(2001)1
\item M.Ausloos et. al, Phys. Rev.B 66(2002)174436
\item E.Dagotto, Nanscale Phase Separation and Clossal Magnetoresistance, Springer-Veralg,Heidelberg 2003
\item T.Becker et. al., Phys. Rev. Lett. 89(2002)237203
\item S.Kumar and P.Majumdar , Phys. Rev.Lett. 92(2004)126602
\item K.H.Ahn et. al., Nature (London) 428(2004)402
\item M.S.Reis et. al., Europhys.Lett. 58(2002)42
\item M.S.Reis et. al., Phys. Rev. B 66(2002)134417
\item M.S.Reis et. al., Phys. Rev. B 68(2003)014404
\item M.S.Reis et. al., Phys. Rev. B73(2006)092401
\item V.G.Bhide,  Mossbauer Effect and its Applications, TATA McGraw Hill Publishing, New Delhi 1973
\item R.S.Preston et. al., Phys. Rev. 128(1962)2207
\item J.L.Dormann and D.Fiorani(editors) , Manetic Properties of Fine Particles , North Holland 1992
\item B.V.Thosar et. al. (editors), Advances in Mossbauer Spectroscopy, Elsevier Publishers 1983
\item R.L.Cohen, Elements of Mossbauer Spectroscopy, Academic Press 1976
\end{enumerate}
\end{document}